\documentclass[journal,12pt, onecolumn]{IEEEtran}
\usepackage{algorithm,algorithmic,amsbsy,amsmath,amssymb,epsfig,bbm,mathrsfs,multirow,amsthm}
\usepackage{bm}
\usepackage{color}
\usepackage{hyperref}
\usepackage{setspace}

\begin{document}

\title{IRS-based Wireless Jamming Attacks:\\ When Jammers can Attack without Power}
\author{\IEEEauthorblockN{Bin Lyu, Dinh Thai Hoang, Shimin Gong, Dusit Niyato, and Dong In Kim \vspace*{-4mm}}
\thanks{B.~Lyu is with Nanjing University of Posts and Telecommunications, China (e-mail: blyu@njupt.edu.cn).}
\thanks{D.~T.~Hoang is with University of Technology Sydney, Australia (email:hoang.dinh@uts.edu.au).}
\thanks{S.~Gong is with Sun Yat-sen University, China (email: gong0012@e.ntu.edu.sg).}
\thanks{D.~Niyato is with Nanyang Technological University, Singapore (email: dniyato@ntu.edu.sg)}
\thanks{D.~I.~Kim is with Sungkyunkwan University, South Korea (e-mail: dikim@skku.ac.kr)}
}

\maketitle

\begin{abstract}
This paper proposes to use Intelligent Reflecting Surface (IRS) as a green jammer to attack a legitimate communication without using any internal energy to generate jamming signals. In particular, the IRS is used to intelligently  reflect the signals from  the legitimate transmitter to  the legitimate receiver (LR) to guarantee that the received signals from direct and reflecting links can be added destructively, which thus diminishes the Signal-to-Interference-plus-Noise Ratio (SINR) at the LR. To minimize the received signal power  at the LR, we consider the joint optimization of  magnitudes of reflection coefficients and discrete phase shifts at the IRS. Based on the  block coordinate descent, semidefinite relaxation, and Gaussian randomization techniques, the solution can be obtained efficiently.  Through simulation results, we show that by using the IRS-based jammer, we can reduce the signal power received at the LR by up to 99\%. Interestingly, the performance of the proposed IRS-based jammer is even better than that of the conventional active jamming attacks in some scenarios.
\end{abstract}
\begin{IEEEkeywords}
Jamming attack, IRS, intelligent backscatter, IoT, low-power sensor networks.
\end{IEEEkeywords}

\section{Introduction}

Recently, intelligent reflecting surface (IRS) has been emerging to be a breakthrough technology which enables significantly improving spectrum and energy efficiency for future wireless communication systems~\cite{Renzo2019Smart}-\cite{Huang2019Survey}. Unlike the conventional backscattering communication techniques which uncontrollably scatter reflected radio frequency (RF) signals~\cite{Huynh2017Survey}, the IRS provides a smart radio environment in which the reflected signals will be controlled in order to remarkably enhance the network performance \cite{Wu2020}. In particular, IRS is a reconfigurable metasurface composed of multiple low-cost passive reflecting elements. Each element can reflect the incident signal independently with an adjustable amplitude
and phase shift controlled by an IRS micro-controller (as illustrated in Fig.~\ref{fig_System_Model}). In this way, by simultaneously adjusting the phase shifts of all elements, the IRS can fully control the strength and direction of the reflected electromagnetic waves, thereby the signal power received at the target devices can be significantly improved. As a result, IRS has a huge potential to fundamentally change how wireless networks are designed and pave the wave for future wireless communication systems.

In the literature, the IRS has been extensively investigated to improve the performance of wireless communication systems  \cite{Wu2019Intelligent}-\cite{QingqingWuAN}. In~\cite{Wu2019Intelligent}, the authors proposed to use an IRS to minimize the total transmit power at the access point (AP) subject to the signal-to-interference-plus-noise ratio (SINR) constraints at the users, for which the active beamforming at the AP and continuous phase shifts at the IRS are jointly optimized.  In \cite{Huang2019Reconfigurable}, the IRS was employed to enhance the system energy efficiency, where the phase shifts are considered to be discrete for practical implementation.
In~\cite{Yu2019Enabling} and \cite{Cui2019Secure}, the authors proposed the idea of using IRS to maximize the achievable secrecy rate to protect the system from eavesdropping attacks. In \cite{Chu2020Secure}, the IRS was utilized to minimize the transmit power at the base station subject to the secrecy rate constraint. Different from \cite{Yu2019Enabling}-\cite{Chu2020Secure} those focused on the joint optimization of transmit beamforming at the base station and reflect beamforming at the IRS, the transmit beamforming with artificial noise and reflect beamforming were further considered in \cite{QingqingWuAN} to enhance the system secrecy rate. It should be noted that in \cite{Wu2019Intelligent}-\cite{QingqingWuAN}, only the design of phase shifts was considered and the reflection coefficients were  set to be one directly.

As mentioned above, most of current research works focus on using the IRS to enhance system performance. However,  the deployment of IRS can also cause  harmful interference to critical wireless communications, which has not been addressed effectively yet. To raise a concern about the proper management of IRS, we introduce an adverse application of IRS in wireless networks, where an IRS is used as a \emph{green jammer} to attack the communication between two legitimate devices. Different from  \cite{Yu2019Enabling}-\cite{QingqingWuAN}, in which the IRS is used to enhance the secrecy rate at the legitimate receiver (LR), we propose to use the IRS-based jammer to degrade the SINR at the LR. In addition, different from conventional active jamming attacks which use their own internal energy to transmit strong noise signals to the victim system, our proposed IRS-based jammer can use directly the signals of the victim communication system to attack by changing their reflection coefficients and phase shifts.
Consequently, the IRS-based jamming attacks can interfere the system without leaving any \emph{energy footprint}, and thus it is very  difficult to be detected and prevented.

In particular, in the system under considerations, once receiving signals transmitted from the legitimate transmitter (LT), the IRS-based jammer will use directly the LT's signals to attack the LR. Specifically, the IRS-based jammer will jointly optimize the magnitudes of reflection coefficients and discrete phase shifts to minimize the total received signal power at the LR.
To address the non-convex problem in jointly determining the optimal magnitudes of reflection coefficients and phase shifts for the jammer, we adopt the block coordinate descent (BCD) method to solve two sub-problems iteratively in an alternating manner, in which the magnitudes of reflection coefficients and phase shifts are optimized respectively.  For the first sub-problem, we first relax the discrete phase shifts to their continuous counterparts, then solve the relaxed sub-problem by using the semidefinite relaxation (SDR) technique \cite{Luo} and the Gaussian randomization method, and finally obtain the solution by quantization. For the second problem, we use the CVX tool \cite{BoydTwo} to solve it efficiently. Through simulation results, we show that our proposed IRS-based jammer can degrade the received signal power at the LR by up to 99\%. More interestingly, we show that in some scenarios, the performance obtained by our proposed IRS-based jammer can be even better than that of conventional active jamming attacks.

\section{System Model}
\label{Sec.System}

\begin{figure}[!]
	\centering
	\includegraphics[scale=0.5]{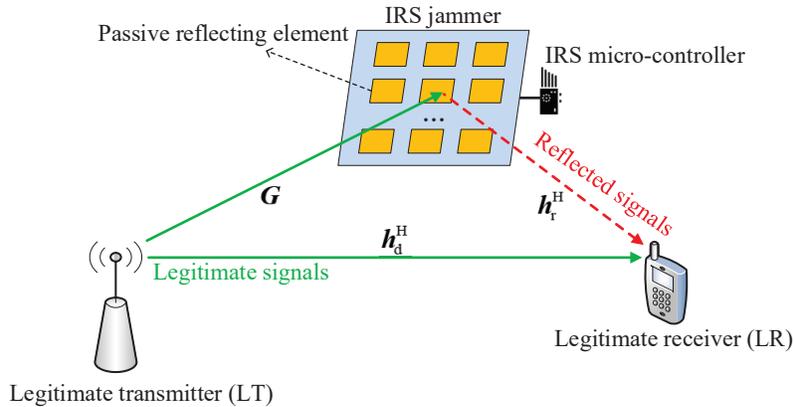}
	\caption{System model.}
	\label{fig_System_Model}
\end{figure}

In this paper, we consider a conventional legitimate communication system including an LT and an LR as illustrated in Fig.~\ref{fig_System_Model}. The LT is equipped with $M$ antennas, while the LR has a single antenna. There is an IRS-based jammer located between the LT and LR with the aim to disturb the legitimate communication. The IRS is equipped with $N$ passive reflecting elements, each of which is equipped with adjustable amplitude and phase shift  to reflect the incident signals independently \cite{Wu2020}. The IRS uses a micro-controller to perform necessary calculation and control functions, such as channel estimation or controlling the switching circuit in passive elements to change the amplitude and phase of the reflected signal. The channels between the LT-LR link, IRS-LR link, and LT-IRS link are denoted by $\bm{h}_\textrm{d}^\textrm{H} \in \mathbb{C}^{1 \times M}$, $\bm{h}_\textrm{r}^\textrm{H} \in \mathbb{C}^{1 \times N}$, and $\bm{G} \in \mathbb{C}^{N \times M}$, respectively, where the superscript $\textrm{H}$ represents the conjugate transpose operation and $\mathbb{C}$ denotes the space of $a \times b$ complex-valued matrices.

 Let  $\bm{\Theta} = \textrm{diag} (\beta_1 e^{j \theta_1}, \ldots, \beta_N  e^{j \theta_N})$  as the IRS's diagonal matrix, where  $\beta_n \in [0, 1]$ and $\theta_n \in [0, 2\pi)$ are the  magnitude of reflection coefficient and phase shift on the combined incident signal, respectively.  For practical implementation of the IRS \cite{Wu2020,Huang2019Reconfigurable}, we consider that the phase shift for each element can only be selected from a set of discrete values, which are assumed to be obtained by uniformly quantizing the interval $[0, 2 \pi)$ for simplicity \cite{QingqingWuDiscrete}. The set of discrete values  is then expressed as:
\begin{align}
\label{PhaseShiftCon}
\theta_n \in \mathcal{F} = \{ 0, 2\pi/L, \ldots, 2 \pi (L-1)/L \},
\end{align}
where $L=2^b$ is the total number of phase shift levels, and $b$ indicates the phase resolution in number of bits \cite{Huang2018Conf}. In this paper, we consider a linear beamforming at the LT with $\bm{\omega} \in \mathbb{C}^{M \times 1}$ denoting the transmit beamforming vector and satisfying $\left\|\boldsymbol{\omega}\right\|^2 = P_\textrm{T}$, where $P_\textrm{T}$ is the total transmit power at the LT. In addition, we only consider the signals reflected by the IRS for the first time at the LR, while those reflected by the IRS  two and more times  are ignored due to the substantial path-loss \cite{Wu2019Intelligent}.
Then, the total signal received at the LR can be expressed as:
\begin{equation}
y = (\bm{h}_\textrm{r}^\textrm{H} \bm{\Theta} \bm{G} + \bm{h}_\textrm{d}^\textrm{H}) \bm{\omega} s + z = (\bm{h}_\textrm{r}^\textrm{H} \bm{\Gamma} \bar{\bm{\Theta}} \bm{G} + \bm{h}_\textrm{d}^\textrm{H}) \bm{\omega} s + z,
\end{equation}
where  $s$ the information-carrying signal with unit power, $z$ denotes the additive white Gaussian noise (AWGN) at the LR with zero mean and variance $\sigma^2$, $\bm{\Gamma} = \textrm{diag} (\beta_1,\ldots,\beta_N)$, and $\bar{\bm{\Theta}} = \textrm{diag} (e^{j \theta_1}, \ldots, e^{j \theta_N}) $. Accordingly, the received signal power  at the LR is given by:
\begin{equation}
\label{RecePower}
\gamma = |(\bm{h}_\textrm{r}^\textrm{H} \bm{\Gamma} \bar{\bm{\Theta}} \bm{G} + \bm{h}_\textrm{d}^\textrm{H}) \boldsymbol{\omega}|^2.
\end{equation}


\section{Received signal power minimization}
\subsection{Problem Formulation}
In this paper, we focus on minimizing  the received signal power in  \eqref{RecePower} by jointly optimizing the  magnitudes of reflection coefficients and  phase shifts.\footnote{In this paper, we consider the ideal phase shift model for simplicity, i.e., the magnitude of reflection coefficient at each reflecting element is independent  of its phase shift, which was widely used in the literature, e.g., \cite{Yu2019Enabling}-\cite{QingqingWuAN}. In the future work, the practical phase shift model \cite{QingqingWuPracticalPhase} can be considered to  investigate the effect of nonlinear dependence between phase shift and reflection amplitude on the attacking performance.}  
Denote $\bm{\beta} = [\beta_1,\ldots, \beta_N]^\textrm{H}$ and $\bm{\theta} = [\theta_1, \ldots, \theta_N]^\textrm{T}$, where the superscript $\textrm{T}$ represents the transpose operation. The corresponding optimization problem can be formulated as follows:
\begin{equation}\tag{$\textbf{P1}$} 
\begin{aligned}
 \min_{\bm{\beta},\bm{\theta}} & |(\bm{h}_\textrm{r}^\textrm{H} \bm{\Gamma}  \bar{\bm{\Theta}} \bm{G} + \bm{h}_\textrm{d}^\textrm{H} ) \bm{\omega} |^2 , \\ 
\text{s.t.}~~ & 0 \le \beta_n \le 1, ~~\forall n=\{1,\ldots,N\}, \\ 
& \theta_n \in \mathcal{F}, ~~\forall n=\{1,\ldots,N\}. 
\end{aligned}
\end{equation}
Note that the objective function of \textbf{P1} is non-convex as $\bm{\beta}$ and $\bm{\theta}$ are coupled. Moreover, $\theta_n$ is restricted to be a discrete value. Hence, \textbf{P1} is a mixed-integer non-linear program (MINLP), which is typically NP-hard and  difficult to solve directly. 

\subsection{Proposed Solution}
To solve \textbf{P1} efficiently, we  propose an algorithm based on the BCD method to optimize $\bm{\beta}$ and $\bm{\theta}$ alternatively. 

\subsubsection{Phase shift design}
Given $\bm{\beta}$, we first design the phase shifts by solving the following problem:
\begin{equation}\tag{$\textbf{P2}$} 
\begin{aligned}
\min_{ \bm{\theta}}&  |(\bm{h}_\textrm{r}^\textrm{H} \bm{\Gamma}  \bar{\bm{\Theta}} \bm{G} + \bm{h}_\textrm{d}^\textrm{H} ) \bm{\omega} |^2 , \\ 
\text{s.t.}~~ & \theta_n \in \mathcal{F}, ~~\forall n=\{1,\ldots,N\}. 
\end{aligned}
\end{equation}
Note that \textbf{P2} is still non-convex as its constraints are non-convex. In order to solve \textbf{P2}, we relax the discrete variables to their continuous counterparts, i.e., $0\le \theta_n \le 2 \pi, ~\forall n$. Then, \textbf{P2} is recast as follows:
\begin{equation}\tag{$\textbf{P2.1}$} 
\begin{aligned}
 \min_{ \bm{\theta}} &|(\bm{h}_\textrm{r}^\textrm{H} \bm{\Gamma}  \bar{\bm{\Theta}} \bm{G} + \bm{h}_\textrm{d}^\textrm{H} ) \bm{\omega} |^2 , \\ 
\text{s.t.}~~ & 0 \le \theta_n \le 2 \pi, ~~\forall n=\{1,\ldots,N\}. 
\end{aligned}
\end{equation}
Let ${v}_{n} =  e^{j \theta_n}$ and $\bm{v} = [v_1,\ldots,  v_N]^\textrm{H}$, where $|v_n| =1, \forall n$. Denote $\bm{\alpha} = \text{diag} (\bm{h}_\textrm{r}^\textrm{H} \bm{\Gamma}) \bm{G}\bm{\omega} $ and $\psi = \bm{h}_\textrm{d}^\textrm{H} \bm{\omega}$. Then, the objective function of \textbf{P2.1} can be  reformulated as $| \bm{v}^\textrm{H} \bm{\alpha} + \psi|^2 =  \bm{v}^\textrm{H} \bm{\alpha} \bm{\alpha}^\textrm{H} \bm{v} + \bm{v}^\textrm{H} \bm{\alpha} \psi^\textrm{H} + \psi \bm{\alpha}^\textrm{H} \bm{v} + |\psi|^2$.  \textbf{P2.1} is thus equivalent to \textbf{P2.2}, which is given by: 
\begin{equation}\tag{$\textbf{P2.2}$} 
\begin{aligned}
\min_{\bm{v}} ~~&  \bm{v}^\textrm{H} \bm{\alpha} \bm{\alpha}^\textrm{H} \bm{v} +\bm{v}^\textrm{H} \bm{\alpha} \psi^\textrm{H} +  \psi \bm{\alpha}^\textrm{H} \bm{v} + |\psi|^2, \\ 
\text{s.t.}~~ & |v_n| = 1, ~~\forall n=\{1,\ldots,N\}.
\end{aligned}
\end{equation}

Note that \textbf{P2.2} is still non-convex. To address this issue, an auxiliary matrix $\bm{R}_i$ and an auxiliary vector $\bm{\mu}$ are further introduced for substitutions, which are expressed as follows:
$$\bm{R} = 
\begin{bmatrix}
 \bm{\alpha} \bm{\alpha}^\textrm{H} &  \bm{\alpha} \psi^\textrm{H} \\ 
 \bm{\alpha}^\textrm{H} \psi & 0 
\end{bmatrix},  \quad 
{\bm{\mu} } = 
\begin{bmatrix}
\bm{v} \\ 
1
\end{bmatrix}.
$$

With $\bm{R}$ and $\bm{\mu}$, the objective function of \textbf{P2.2} can be recast by: 
$\text{Tr}(\bm{R} \bm{\mu} \bm{\mu}^\textrm{H}) + |\psi|^2$. Let $\bm{V} = \bm{\mu} \bm{\mu}^\textrm{H}$, where $\bm{V} \succeq 0 $ and $\text{rank} (\bm{V}) =1$. Then, $\text{Tr}(\bm{R} \bm{\mu} \bm{\mu}^\textrm{H}) + |\psi|^2$ is equivalent to  $\text{Tr} (\bm{R} \bm{V}) + |\psi|^2$.
In addition, we have a new constraint that $\bm{V}_{n,n} = 1$ due to that $|v_n| = 1,~\forall n$, where $\bm{V}_{n,n}$ denotes the $n$-th diagonal
element of $\bm{V}$. Hence, \textbf{P2.2} can be reformulated as: 
\begin{equation}\tag{$\textbf{P2.3}$} 
\begin{aligned}
\min_{\bm{V}} ~~& \text{Tr} (\bm{R} \bm{V}) + |\psi|^2, \\ 
\text{s.t.}~~ & \bm{V}_{n,n} = 1,~ \forall n, \\ 
& \bm{V} \succeq 0, \\ 
& \text{rank} (\bm{V}) = 1.
\end{aligned}
\end{equation}
\textbf{P2.3} is still a non-convex optimization problem due to the rank-one constraint. However, we can apply the SDR technique \cite{Luo} to  relax the rank-one constraint. After that, \textbf{P2.3} is recast as:
\begin{equation}\tag{$\textbf{P2.4}$} 
\begin{aligned}
\min_{\bm{V}} ~~& \text{Tr} (\bm{R} \bm{V}) + |\psi|^2, \\ 
\text{s.t.}~~ & \bm{V}_{n,n} = 1, ~\forall n, \\ 
& \bm{V} \succeq 0.
\end{aligned}
\end{equation}
It is obvious that \textbf{P2.4} is a convex SDP \cite{BoydOne} and can be thus solved by the CVX tool \cite{BoydTwo}. However, the solution obtained from \textbf{P2.4} solved by the CVX is generally not a rank-one solution, which leads to a lower-bound of the received signal power at the LR. Thus, we propose to utilize the Gaussian randomization method to construct an approximate solution for \textbf{P2.3} based on the solution from \textbf{P2.4}. Denote the solution for \textbf{P2.4} as $\hat{\bm{V}}$, its singular value decomposition is given by: 
\begin{align}
\hat{\bm{V}} = \bm{U} \bm{\varSigma} \bm{U}^\textrm{H},
\end{align}
where $\bm{U} \in \mathbb{C}^{(N+1) \times (N+1)}$ and $\bm{\varSigma} \in \mathbb{C}^{(N+1) \times (N+1)}$ are the unitary matrix and the diagonal matrix of $\hat{\bm{V}}$, respectively. Based on the Gaussian randomization method, a rank-one solution for \textbf{P2.3} can be constructed by $\bar{\bm{V}} = \bar{\bm{\mu}} \bar{\bm{\mu}}^\textrm{H},$
where 
\begin{align}
\label{BeamRandom}
\bar{\bm{\mu}} = \bm{U} \sqrt{\bm{\varSigma}} \bm{r}, 
\end{align}
and the random vector $\bm{r}\in \mathbb{C}^{(N+1)\times 1} $ is independently generated according to $\mathcal{CN}(0, \bm{I}_{N+1})$. With $\bar{\bm{\mu} }$, an approximate solution for \textbf{P2.1}, denoted by $\bar{\bm{\theta}}$, is recovered by:
\begin{align}
\label{OptiTheta}
\bar{\bm{\theta}} = - \arg \Big( \Big[\frac{\bar{\bm{\mu}}}{\bar{\bm{\mu}}(N+1)}\Big]_{(1:N) }\Big),
\end{align}
where $[\bm{w}]_{(1:N)}$ indicates the first $N$ elements selected from the vector $\bm{w}$,  $\arg(\bm{w})$ denotes the phase of each element in the vector $\bm{w}$, and $\bar{\bm{\mu}} (N+1)$ is the $(N+1)$-th element of $\bar{\bm{\mu}}$. Then, we quantize $\bar{\bm{\theta}}$ to their nearest discrete values  in $\mathcal{F}$ defined in \eqref{PhaseShiftCon}, which is denoted by $\hat{\bm{\theta}}$.

Through generating $\bm{r}$ for a sufficiently large number, we can find the best
$\hat{\bm{\theta}}$ among all $\bm{r}$'s that minimizes the received signal power at the LR, which is considered as the sub-optimal solution for \textbf{P2} with high accuracy \cite{SPR}. The algorithm for solving \textbf{P2} is summarized in Algorithm \ref{Alg:One}.


\begin{algorithm}
	\caption{ The Algorithm for solving \textbf{P2}.}
	\label{Alg:One}
	\begin{algorithmic}[1] 
		\STATE{Initialize $D$, which is a sufficiently large number of generating random vectors.  }
		\STATE Solve \textbf{P2.4}   and obtain $\hat{\bm{V}}$.
		\STATE Compute the singular value decomposition of $\hat{\bm{V}}$, and obtain $\bm{U} $ and $\bm{\varSigma}$. 
		\STATE Initialize $\Gamma = \emptyset$.
		\FOR{$d=1,\ldots,D$}
		\STATE {Generate $\bm{r}$ according to $\mathcal{CN}(0, \bm{I}_{N+1})$. }
		\STATE {Construct $\bar{\bm{\mu}}$ by \eqref{BeamRandom}, and obtain $\bar{\bm{\theta}}$ by \eqref{OptiTheta}. }
		\STATE{Quantize $\bar{\bm{\theta}}$ to their nearest discrete values in $\mathcal{F}$ and obtain $\hat{\bm{\theta}}$.}
		\STATE {Compute the objective function value of \textbf{P2} with $\hat{\bm{\theta}}$, which is denoted by $\gamma_d$.}
		\STATE {$\Gamma = {\Gamma} \cup \gamma_d$. }
		\ENDFOR 
		\STATE{ Obtain the sub-optimal solution for \textbf{P2} by ${\bm{\theta}}^*  = \arg \min_{\hat{\bm{\theta}} } \Gamma $. }
	\end{algorithmic}
\end{algorithm}

\subsubsection{Reflection coefficient design}
Given $\bm{\theta}$, we proceed to design the magnitudes of reflection coefficients by solving \textbf{P3}, which is given by:
\begin{equation}\tag{$\textbf{P3}$} 
\begin{aligned}
\min_{ \bm{\beta}} ~~&  |(\bm{h}_\textrm{r}^\textrm{H} \bm{\Gamma}  \bar{\bm{\Theta}} \bm{G} + \bm{h}_\textrm{d}^\textrm{H} ) \bm{\omega} |^2 , \\ 
\text{s.t.}~~ & 0 \le \beta_n \le 1, ~~\forall n= \{1,\ldots,N\}. 
\end{aligned}
\end{equation}
Similarly, \textbf{P3} can be reformulated as follows:
\begin{equation}\tag{$\textbf{P3.1}$} 
\begin{aligned}
 \min_{ \bm{\beta}} ~~& |\bm{\beta}^H \bm{c} + \psi |^2 , \\ 
\text{s.t.}~~ & 0 \le \beta_n \le 1, ~~\forall n= \{1,\ldots,N\},
\end{aligned}
\end{equation}
where $\bm{c} = \textrm{diag} (\bm{h}_r^H) \bar{\bm{\Theta}}\bm{G}\bm{\omega} $.
 It is obvious that \textbf{P3.1} is the convex optimization problem. Hence, we use the CVX tool to solve it directly.

Based on the above analysis, the algorithm to solve \textbf{P1} is summarized in Algorithm \ref{Alg:Two}.\footnote{In Section \ref{Simulation}, we sequentially solve \textbf{P2} and \textbf{P3.1}  for only one time, which can still achieve satisfying performance.} 
Due to that the received signal power is always non-increasing after solving \textbf{P2} and  \textbf{P3.1} sequentially in each iteration, Algorithm \ref{Alg:Two} can converge to the given threshold.

\begin{algorithm}
	\caption{ The Algorithm for solving \textbf{P1}.}
	\label{Alg:Two}
	\begin{algorithmic}[1] 
		\STATE{\text{Initialize}: $i=0$ and $\bm{\beta}^{(0)} = [\beta_1^{(0)}, \ldots, \beta_N^{(0)}]^\textrm{H}$.  }
		\REPEAT
		\STATE{Solve \textbf{P2} via Algorithm \ref{Alg:One} for the given $\bm{\beta}^{(i)}$, and obtain the quantized phase shifts ${\bm{\theta}}^{*(i+1)}$.}
		\STATE{Obtain $\bm{\beta}^{(i+1)}$ by solving \textbf{P3.1} for the given ${\bm{\theta}}^{*(i+1)}$. }
		\STATE{$i=i+1$.}
		\UNTIL {The convergence is achieved.}

	\end{algorithmic}
\end{algorithm}


\section{Performance Evaluation}
\label{Simulation}
The simulated network topology is a 2D coordinate system, where the coordinates of the LT, the LR, and the IRS are given as ($x_t$,0), $(x_r,0)$, $(x_i,y_i)$, respectively. We follow the channel model considered in \cite{Chu2020Secure}. The small-scale channel coefficients are generated to be circularly complex Gaussian random variables with zero mean and unit variance. The large-scale path-loss is modeled as $A (d/{d_0})^{-\alpha}$, where $A$ is the path-loss at the reference distance $d_0=1$ m and set to $A=-30$ dB, $d$ denotes the distance between two nodes, and $\alpha$ is the path-loss exponent. The path-loss exponents of the LT-LR link, the LT-LRS link, and the LRS-LR link are set to $\alpha_{\text{LT-LR}} = 3.5$, $\alpha_{\text{LT-IRS}} = 2.8$, and $\alpha_{\text{IRS-LR}} = 2.8$, respectively \cite{Wu2019Intelligent}.
Similar as \cite{Chu2020Secure}, the setting of other parameters is given as follows unless otherwise stated: $P_T = 30$ dBm, $M=8$, $N=150$, $b=5$, $x_t=0$, $x_r=10$ m, $x_i=5$ m, and $y_i=2$ m. The traditional active jamming scheme with transmit power $P_\textrm{a}$ is used as a benchmark, where the active jammer is deployed at the location of the IRS. In addition, the scheme without jamming is also used as a benchmark to evaluate the impact of the IRS jamming attack.

\begin{figure}[t]
	\centering
	\includegraphics[scale=0.7]{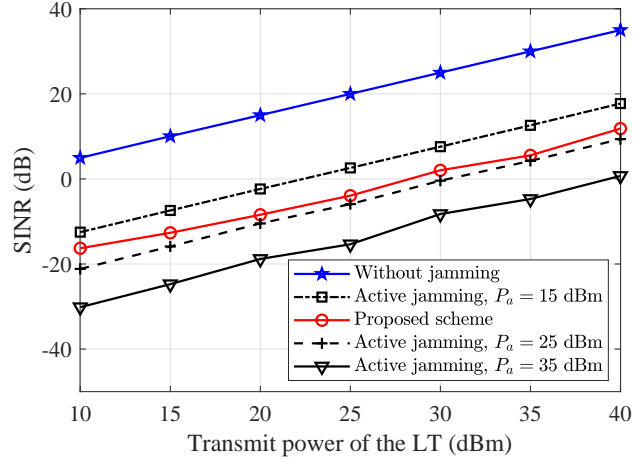}
	\caption{SINR versus transmit power of the LT. }
	\label{SNRvsTransPower}
\end{figure} 

\begin{figure}[t]
	\centering
	\includegraphics[scale=0.7]{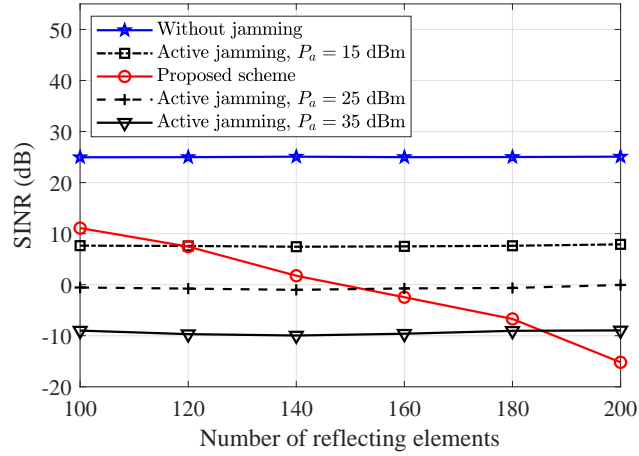}
	\caption{SINR versus number of reflecting elements. }
	\label{SNRvsNumElements}
\end{figure} 

In order to evaluate the performance of the proposed IRS jammer with the conventional active jamming attack, we first vary the LT's transmit power and the number of reflecting elements and compare their performance in terms of SINR. The SINR is determined by:
\begin{equation}
\text{SINR} = \frac{P_{\text{signal}}}{P_{\text{noise}} + P_{\text{a}}} .
\end{equation}
In the case of the conventional active jamming attack, $P_{\text{a}}$ is the power of received jamming signals transmitted by  the active jammer, while $P_{\text{a}}$ is set to be zero in the case of IRS jamming attack. The noise power $P_{\text{noise}}$ in both cases is set at $\sigma^2=-60$ dBm \cite{Chu2020Secure}. 

In Fig.~\ref{SNRvsTransPower}, we vary the LT's transmit power and show the SINR at the LR under different attacks. It can be observed that our proposed scheme can  degrade the SINR at the LR by about 22 dB compared to the scheme without jamming. Also, it can be observed that when we increase the transmit power of the conventional active jammer, the SINR at the LR will be reduced. Here, it is obvious that the active jamming scheme with high jamming power (e.g., $P_a = 35$ dBm) can achieve the lowest SINR, i.e., the best attack performance. However, interestingly, if the active jamming power is not high, i.e., $P_a = 15$ dBm, our proposed IRS jammer can  achieve better attack performance even without using any internal energy to launch jamming attacks. The reason is that in our proposed scheme, the  magnitudes of reflection coefficients and phase shifts of all passive reflecting elements can be carefully designed such that the total received signals at the LR from the direct and reflecting links can be added destructively, which can significantly reduce the received power and thus diminish the SINR at the LR.

In Fig.~\ref{SNRvsNumElements}, we show the influence of number of reflecting elements on the SINR. It can be observed that as the number of reflecting elements increases, the SINR at the LR under the IRS jamming attack will be significantly reduced due to very low signal power received at the LR. Interesting, if the number of reflecting elements is sufficiently large, e.g., 200, our proposed IRS jammer can even achieve a better attack performance than that of the active jamming scheme with $P_\textrm{a} = 35$ dBm. In addition, we can conclude that the attack performance of our proposed scheme can be further enhanced with the increase of reflecting elements, which confirms the superiority of our proposed scheme.

\begin{figure}[t]
	\centering
	\includegraphics[scale=0.7]{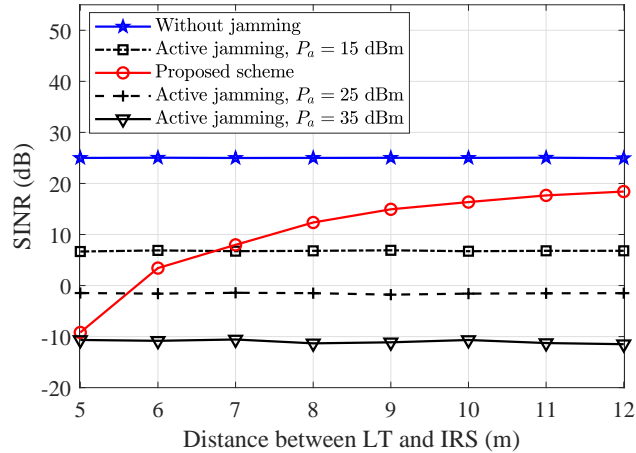}
	\caption{SINR versus distance between LT and IRS.}
	\label{SNRvsDistanceLTandIRS}
\end{figure}

Fig.~\ref{SNRvsDistanceLTandIRS} shows the received SINR versus the distance between the LT and IRS  by changing the IRS's coordinate, where the distance between the LR and IRS is a constant, i.e., 5 m. From Fig.~\ref{SNRvsDistanceLTandIRS}, we can observe that our proposed scheme can always degrade the received SINR at the LR compared to the scheme without jamming. As the distance between the LT and IRS increases from 5 m to 12 m, the received SINR increases.
 If the distance between the LT and IRS is shorter than 7 m, the attack performance of our proposed scheme is better than that of the active jamming scheme with $P_a = 15$ dBm. 
 However, if the distance between the LT and IRS is larger than the threshold, i.e., 7 m, our proposed scheme cannot achieve the performance as good as that of the active jamming scheme. The reason is that as the distance increases, the  signal power received at the LR from the reflecting links reduces, which limits the performance of destructive addition and thus declines the effect of IRS jamming attack.

\begin{figure}[t]
	\centering
	\includegraphics[scale=0.7]{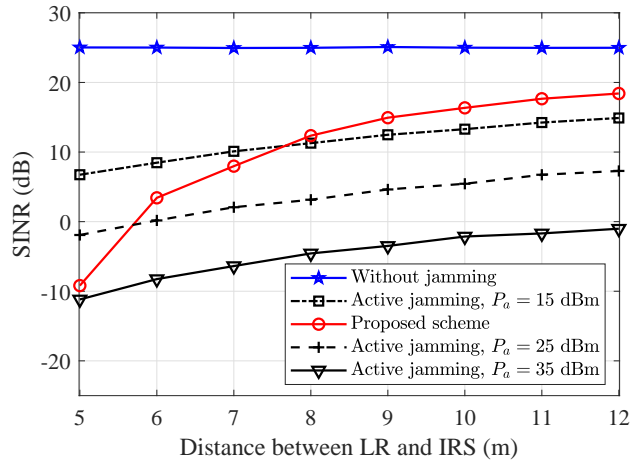}
	\caption{SINR versus distance between LR and IRS.}
	\label{SNRvsDistanceIRSandLR}
\end{figure} 

Fig.~\ref{SNRvsDistanceIRSandLR} investigates the received SINR versus the distance between the LR and IRS  by changing the IRS's coordinate, where the distance between the LT and IRS is a constant, i.e., 5 m.
As the distance between the LR and IRS increases from 5 m to 12 m, the  SINRs at the LR achieved by the proposed scheme and the active jamming scheme increase.  
 If the distance between the LR and IRS is shorter than 8 m, the attack performance of our proposed scheme is also better than that of the active jamming scheme with $P_a = 15$ dBm. In particular, when  the distance is 5 m, the attack performance of  our proposed scheme is even close to that of the active jamming scheme with $P_a = 35$ dBm.
If the distance is larger than the threshold, i.e., 8 m, the attack performance achieved by our proposed scheme is still close to that of the active jamming scheme with $P_a = 15$ dBm.

\section{Conclusions}

In this paper, we introduce a new type of jamming attack, called IRS jamming. Different from all other wireless jamming attacks in the literature which have to use their own energy to generate jamming signals, our proposed IRS jammer can leverage directly the LT's signals to reduce the SINR at the LR. This is due to the fact that the IRS jammer can control its reflection coefficients and phase shifts to reflect signals which can significantly reduce the received signal power at the LR. Through simulation results, we show that our proposed IRS jammer can not only remarkably reduce the LR's SINR, but also achieve better performance than that of conventional active jamming attack in some scenarios.

\end{document}